\documentclass[twocolumn,superscriptaddress,amsmath,amssymb,pra,longbibliography]{revtex4-1}

\usepackage{amssymb,amsmath}
\DeclareMathAlphabet\mathbfcal{OMS}{cmsy}{b}{n}
\usepackage{cmap}
\usepackage{graphicx} 
\usepackage{bm}
\usepackage{color}
\usepackage{blindtext}
\usepackage{hyperref}
\usepackage{listings}
\usepackage{booktabs}
\usepackage{multirow}
\usepackage{amsmath}
\usepackage{mathrsfs}
\usepackage{braket}
\usepackage{mathtools}

\usepackage{dsfont}
\usepackage{tabularx}

\begin{document}

\title{On the Ehrenfest theorem and centroids of relativistic particles}

\author{Konstantin Y. Bliokh}
\affiliation{Donostia International Physics Center (DIPC), Donostia-San Sebasti\'{a}n 20018, Spain}
\affiliation{Theoretical Quantum Physics Laboratory, RIKEN Cluster for Pioneering Research, Wako-shi, Saitama 351-0198, Japan}
\affiliation{Centre of Excellence ENSEMBLE3 Sp. z o.o., 01-919 Warsaw, Poland}


\begin{abstract}
We consider relativistic versions of the Ehrenfest relation between the expectation values of the coordinate and momentum of a quantum particle in free space: $d\langle {\bf r} \rangle /dt = \langle {\bf p} \rangle/m$. We find that the simple proportionality between the mean velocity and momentum holds true only for the simplest quadratic dispersion (i.e., dependence of the energy on the momentum). For relativistic dispersion, the mean velocity is generally not collinear with the mean momentum, but velocity of the {\it energy centroid} is directed along the mean momentum. This is related to the conservation of the Lorentz-boost momentum and has implications in possible decomposition of the mean orbital angular momentum into intrinsic and extrinsic parts. Neglecting spin/polarization effects, these properties depend solely on the dispersion relation, and can be applied to any waves, including classical electromagnetic or acoustic fields.
\end{abstract}

\maketitle

\section{Introduction}

The Ehrenfest theorem, derived in early days of quantum mechanics \cite{Ehrenfest1927, Bohm_book}, showed that the expectation values of the quantum-mechanical position and momentum operators obey dynamical equations similar to the classical-mechanics equations. In this manner, it accompanies the quantum-to-classical transition, where the mean values of the coordinates and wavevectors of a semiclassical wavepacket follow the corresponding classical trajectory. 

The simplest free-space version of the Ehrenfest theorem states:
\begin{equation}
\frac{d\langle {\bf r} \rangle}{dt} = \frac{\langle {\bf p} \rangle}{m}\,,\quad
\frac{d\langle {\bf p} \rangle}{dt} = {\bf 0}\,,
\label{Ehr0}    
\end{equation}
where $m$ is the mass of the particle, ${\bf r}$ is the position, ${\bf p}$ is the momentum, and the normalized expectation values of a quantity $Q$ are calculated as $\langle Q \rangle = \langle \psi | \hat{Q} | \psi \rangle /  \langle \psi | \psi \rangle$ using the corresponding operator $\hat{Q}$ and wavefunction $\psi$. Evidently, Eqs.~(\ref{Ehr0}) correspond to the classical-mechanics equations for a free particle described by the Hamiltonian $H({\bf p}) = p^2/2m$, and these are derived for the corresponding quantum Hamiltonian $\hat{H} = \hat{p}^2/2m$.

Here we consider extensions of the Ehrenfest theorem to relativistic particles in free space. The second equation (\ref{Ehr0}) is the conservation of the total momentum, which remains unchanged in the relativistic case, whereas the first equation (\ref{Ehr0}) raises nontrivial questions. The classical relativistic-particle equation reads \cite{LLfield}
\begin{equation}
\frac{d {\bf r} }{dt}= \frac{\partial E}{\partial {\bf p}} = \frac{c^2 {\bf p} }{E}\,,
\label{Class}    
\end{equation}
where $c$ is the speed of light, $E$ is the energy of the particle, and we used the relativistic dispersion relation $E({\bf p})= \sqrt{p^2 c^2 + m^2 c^4}$. In this paper, we show that there are two `Ehrenfest versions' of Eq.~(\ref{Class}), which are based on different mean positions in the left-hand side: (i) the particle-probability centroid $\langle {\bf r} \rangle$ and (ii) the energy centroid $\langle {\bf r}_E \rangle$. The velocity of the probability centroid is generally not aligned with the mean momentum $\langle {\bf p} \rangle$, while the velocity of the energy centroid is.
 
The ambiguity of the Ehrenfest relation for relativistic quantum particles owes its origin to the relativistic {\it non-quadratic dispersion} $E({\bf p})$. Therefore, this problem is equally relevant to classical electromagnetic and acoustic waves with linear (massless) dispersion, and any other waves with non-quadratic dispersions. 

Note that recent generalization of the Ehrenfest theorem for quantum relativistic particles \cite{Bialynicki-Birula2022} considered only the extension of the second equation (\ref{Ehr0}) in the presence of external fields, and hence does not overlap with our study. Here we  deal with the simplest approximation of a scalar wavefield in free space.  

\section{Relativistic analogues of the Ehrenfest relation}

Our consideration will not involve any spin/polarization effects, and therefore we will consider a scalar wavefunction $\psi$ describing quantum or classical wavefield. This is justified for semiclassical/paraxial wavepackets (large as compared with the central wavelength) with a fixed polarization, which ensures that the spin-orbit coupling effects are negligible. In the same approximation, the problem of the definition of the position operator of a relativistic particle \cite{Bacry_book,Pryce1948,Bliokh2017PRA} becomes inessential and does not affect the expectation values we deal with.  
We will also neglect possible effects of the negative-energy states, which can appear in the Dirac equation, assuming that we deal with a pure positive-energy wave state \cite{Thaller_Dirac}. 

\subsection{The first approach}

Let us start with the normalized expectation value of the position operator using the wavefunction in the momentum (Fourier) representation: 
\begin{equation}
\langle {\bf r} \rangle = \frac{\langle \psi | \hat{\bf r} | \psi \rangle}{\langle \psi | \psi \rangle}
= \frac{\int \tilde{\psi}^* e^{i\omega t} (i{\bm \nabla}_{\bf k}) \tilde{\psi} e^{-i\omega t} \, g d^3{\bf k}}{\int \tilde{\psi}^* \tilde{\psi}\, g d^3{\bf k}} \,.
\label{R}    
\end{equation}
Here ${\bf k}$ is the wavevector, $\hat {\bf r} = i{\bm \nabla}_{\bf k}$ is the position operator in the momentum representation (we use the $\hbar =1$ units), $\tilde{\psi} ({\bf k})$ is the Fourier amplitude of the wavefunction, $\omega ({\bf k})$ is the frequency and the dispersion relation ($\omega ({\bf k}) \equiv E({\bf p})$), and $g(\omega)$ is a coefficient involved in the inner product for a given type of the wavefield.  
For a relativistic particle with spin $s=0,1/2,1,...$, $g \propto \omega^{1 - 2s}$, but this factor is inessential for our consideration and can be incorporated in the definition of the wavefunction: $\tilde{\psi} \to \tilde{\psi}/\sqrt{g}$. 

Taking the time derivative of Eq.~(\ref{R}), we obtain
\begin{equation}
\frac{d\langle {\bf r} \rangle}{dt} 
= \frac{\int \tilde{\psi}^* ({\bm \nabla}_{\bf k} \omega) \tilde{\psi} \, g d^3{\bf k}}{\int \tilde{\psi}^* \tilde{\psi}\, g d^3{\bf k}} \,.
\label{Vg}    
\end{equation}
Since ${\bm \nabla}_{\bf k} \omega = {\bf v}_g$ is the group velocity for a plane wave, the right-hand side of Eq.~(\ref{Vg}) can be interpreted as the expectation value of the group velocity, $\langle {\bf v}_g \rangle$. For the relativistic dispersion $\omega = \sqrt{k^2 c^2 + m^2 c^4}$, we have ${\bm \nabla}_{\bf k} \omega = c^2{\bf k}/\omega$. As $\hat{\bf p} = {\bf k}$ and $\hat{E} = \omega$ are the operators of the momentum and energy in the Fourier representation, we can write Eq.~(\ref{Vg}) as 
\begin{equation}
\frac{d\langle {\bf r} \rangle}{dt} 
= \left\langle \frac{\partial \omega}{\partial {\bf k}}  \right\rangle \equiv \langle {\bf v}_g \rangle 
= \left\langle \frac{c^2{\bf p}}{E}  \right\rangle \,.
\label{Ehr1}    
\end{equation}

Equation~(\ref{Ehr1}) is the first `Ehrenfest version' of Eq.~(\ref{Class}).
Importantly, the expectation value of the momentum-to-energy ratio in this equation is not equal to the ratio of the expectation values: 
\begin{equation}
\left\langle \frac{{\bf p}}{E}  \right\rangle \neq  \frac{\langle {\bf p} \rangle}{\langle E \rangle}  \,.
\label{ratio}    
\end{equation}
Therefore, Eq.~(\ref{Ehr1}) is not proportional to the usual expectation value of the momentum, $\langle {\bf p} \rangle$. This difficulty can be overcome in another approach, which essentially uses relativistic properties.  
  
\subsection{The second approach}

Instead of the expectation value of the position operator, let us consider an {\it energy centroid} of the field:
\begin{equation}
\langle {\bf r}_E \rangle = \frac{\langle \psi | \hat{E} \hat{\bf r} | \psi \rangle}{\langle \psi | \hat{E} | \psi \rangle}
= \frac{\int \tilde{\psi}^* e^{i\omega t} (i\omega {\bm \nabla}_{\bf k}) \tilde{\psi} e^{-i\omega t} \, g d^3{\bf k}}{\int \tilde{\psi}^* \omega \tilde{\psi}\, g d^3{\bf k}} \,.
\label{RE}    
\end{equation}
Differentiating this equation with respect to time, we readily derive, akin to Eqs.~(\ref{Vg}) and (\ref{Ehr1}):  
\begin{equation}
\frac{d\langle {\bf r}_E \rangle}{dt} 
= \frac{\left\langle  \omega\, { \partial \omega}/{\partial {\bf k}} \right\rangle}{\langle \omega \rangle}  =  \frac{c^2\langle {\bf p}\rangle}{\langle E \rangle} \,.
\label{Ehr2}    
\end{equation}

Equation (\ref{Ehr2}) is the second `Ehrenfest version' of Eq.~(\ref{Class}). It does involve the expectation values of the momentum and energy, but only in the case of the relativistic dispersion. The proportionality of the energy-centroid velocity and the mean momentum has a fundamental relativistic origin. It can be derived for a system of classical relativistic particles \cite{LLfield}, as well as within relativistic field theory \cite{Barnett2011JO,Bliokh2013NJP}, from the conservation of the total {\it boost momentum} in the system:  
\begin{equation}
\langle {\bf N} \rangle = \langle t\, {\bf p} c - {\bf r} E/c \rangle = {\bf const}\,,
\label{BM}    
\end{equation}
and using the relation $\langle {\bf r} E \rangle = \langle {\bf r}_E \rangle \langle E \rangle$.
The boost momentum is a spatiotemporal part of the relativistic angular-momentum tensor, and its conservation is a consequence of the symmetry of the system with respect to the Lorentz boosts. It is easy to see that the equation $d \langle {\bf N} \rangle / dt = {\bf 0}$ is equivalent to Eq.~(\ref{Ehr2}). 

Note, however, that for any dispersion differing from the relativistic one, the energy-centroid velocity is not proportional to the mean momentum. Furthermore, the energy-centroid position is not an expectation value of an operator, it is a {\it ratio} of two expectation values: $\langle {\bf r}_E \rangle = \langle {\bf r} E \rangle / \langle E \rangle$. Therefore, in contrast to Eqs.~(\ref{Ehr0}) and (\ref{Ehr1}), Eq.~(\ref{Ehr2}) is an equation for ratios of the expectation values. 

\begin{figure*}[t!]
\includegraphics[width=0.7\linewidth]{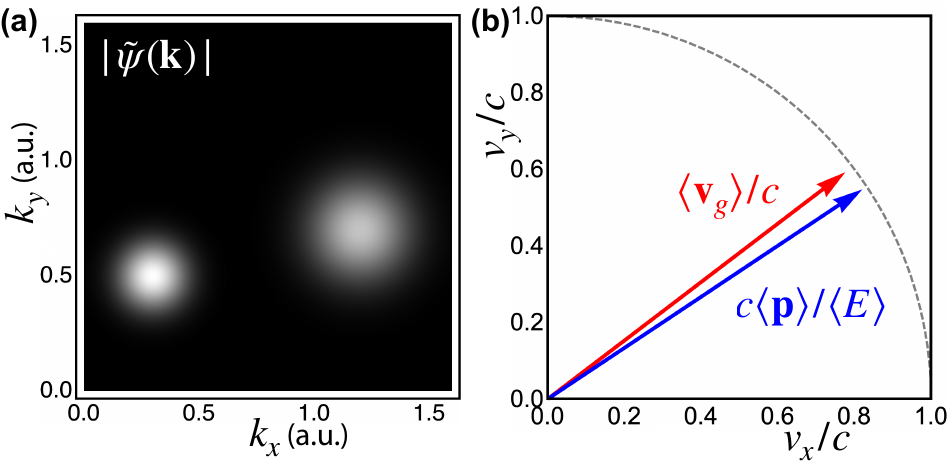}
\caption{(a) A model example of the wavefunction (\ref{example_psi}) distribution in the ${\bf k}$-space. The parameters are: $A_1 = 0.5$, $A_2 = 0.87$, ${\bf k}_1 = (0.3,0.5,0)$, ${\bf k}_2 = (1.2,0.7,0)$, $\Delta_1 = 0.1$, $\Delta_2 = 0.15$ (in arbitrary units). (b) The corresponding velocities of the particle-centroid (red) and energy-centroid (blue) motions, Eqs.~(\ref{example_1}) and (\ref{example_2}), in the case of massless dispersion $\omega = kc$. The dashed curve corresponds to $|{\bf v}|/c=1$. \label{Fig1}}
\end{figure*}
    
\subsection{Comparison and example}

Both equations (\ref{Ehr1}) and (\ref{Ehr2}) describe rectilinear uniform motions of the particle (probability) and energy centroids in free space, because the right-hand sides of these equations are time-independent. However, directions and magnitudes of the corresponding velocities do not generally coincide due to Eq.~(\ref{ratio}). One can say that {\it the mean group velocity is not collinear with the mean momentum} in the general case: $\langle {\bf v}_g \rangle \nparallel \langle {\bf p} \rangle$. These vectors are always collinear only in the case of quadratic dispersion: $\omega = k^2/2m$, ${\bf v}_g = {\bf k}/m$, which reduces Eq.~(\ref{Ehr1}) to the non-relativistic Ehrenfest equation (\ref{Ehr0}).

For a model example, let us consider a wavefield given by a superposition of two Gaussian wavepackets centered around wavevectors ${\bf k}_1$ and ${\bf k}_2$:
\begin{equation}
\frac{\tilde{\psi}}{\sqrt{g}} = \frac{A_1}{\Delta_1^3 \pi^{\frac{3}{2}}} \exp\!\left[ -\frac{({\bf k} - {\bf k}_1)^2}{2\Delta_1^2} \right] + \frac{A_2}{\Delta_2^3 \pi^{\frac{3}{2}}} \exp\!\left[ -\frac{({\bf k} - {\bf k}_2)^2}{2\Delta_2^2} \right],
\label{example_psi}    
\end{equation}
where $A_{1,2}$ are constant amplitudes, normalized as $|A_1|^2 + |A_2|^2 =1$, and $\Delta_{1,2}$ are the widths of the wavepackets in momentum space. Assuming that $|{\bf k}_1 - {\bf k}_2| \gg \Delta_{1,2}$, the overlap between the wavepackets is exponentially small, and we can neglect the cross-terms in the expectation values.     
In doing so, we obtain: 
\begin{equation}
\langle \psi | \psi \rangle \simeq |A_1|^2 + |A_2|^2 =1\,.
\label{example_norm}    
\end{equation}
Assuming also wavepackets to be well-localized away from the origin, $|{\bf k}_{1,2}| \gg \Delta_{1,2}$, the expectation values of an operator $\hat{Q} = Q ({\bf k})$, can be approximated as
\begin{equation}
\langle \psi | \hat{Q} |\psi \rangle \simeq {Q({\bf k}_1) |A_1|^2 + Q({\bf k}_2)|A_2|^2}\,.
\label{example_Q}    
\end{equation}

In this approximation, assuming the relativistic dispersion, we obtain for Eqs.~(\ref{Ehr1}) and (\ref{Ehr2}):
\begin{align}
\label{example_1}
\frac{d\langle {\bf r} \rangle}{dt}  \simeq c^2 \!\left( \frac{{\bf k}_1}{\omega_1} |A_1|^2 + \frac{{\bf k}_2}{\omega_2} |A_2|^2 \right), \\
\frac{d\langle {\bf r}_E \rangle}{dt}  \simeq c^2\, \frac{{\bf k}_1 |A_1|^2 + {\bf k}_2 |A_2|^2}{\omega_1 |A_1|^2 + \omega_2 |A_2|^2}\,,
\label{example_2}    
\end{align}
where $\omega_{1,2} = \omega ({\bf k}_{1,2})$. The velocities in Eqs.~(\ref{example_1}) and (\ref{example_2}) generally have different magnitudes and directions, as shown in Fig.~\ref{Fig1}.      

\section{Relation to the angular momentum}

Position of a particle is important for its angular momentum, which is a fundamental conserved quantity. For a scalar field, with no spin angular momentum, the expectation value of the angular momentum is $\langle {\bf L} \rangle = \langle \psi | \hat{\bf L} | \psi \rangle$, where $ \hat{\bf L} =  \hat{\bf r} \times  \hat{\bf p} = - i {\bf k} \times {\bm \nabla}_{\bf k}$ is the orbital angular momentum operator in the momentum representation. 

In non-relativistic mechanics, the total angular momentum of a distributed system can be decomposed into `intrinsic' and `extrinsic' parts describing the angular momentum with respect to the system's centroid and the angular momentum of the motion of the system `as a whole' \cite{LL_mech}:
\begin{equation}
\langle {\bf L} \rangle^{\rm ext} = \langle {\bf r} \rangle \times \langle {\bf p} \rangle\,, \quad
\langle {\bf L} \rangle^{\rm int} = \langle {\bf L} \rangle - \langle {\bf L} \rangle^{\rm ext} \,.
\label{AM_class}    
\end{equation}
It is easy to show, using Eqs.~(\ref{Ehr0}), that both the extrinsic and intrinsic angular momenta (\ref{AM_class}) are separately conserved quantities.

In the relativistic case, there are two centroids, $\langle {\bf r} \rangle$ and $\langle {\bf r}_E \rangle$, which can be used for the separation of the intrinsic and extrinsic angular momenta. First, it should be noted that neither of these is a Lorentz-covariant quantity (e.g., a Lorentz boost can induce transverse shifts of the centroids \cite{Bliokh2012PRL, Smirnova2018}), so the intrinsic and extrinsic angular momenta are not Lorentz-covariant either. Only the Poincar\'{e} invariants $\langle E \rangle$, $\langle {\bf p} \rangle$, $\langle {\bf L} \rangle$, and $\langle {\bf N} \rangle$ are the Lorentz-covariant conserved quantities. 

Nonetheless, the intrinsic-extrinsic decomposition of the angular momentum can be useful in a given reference frame. On the one hand, using the decomposition (\ref{AM_class}) with the probability centroid $\langle {\bf r} \rangle$ is intuitively clear and helps to solve some paradoxes in the calculations of the angular momenta \cite{Bliokh2012PRL, Smirnova2018, Bliokh2023PRA}. However, such extrinsic and intrinsic angular momenta are generally not conserved in free space \cite{Porras2024}:
\begin{equation}
\frac{d \langle {\bf L} \rangle^{\rm ext}}{dt} = \frac{d \langle {\bf r} \rangle}{dt} \times \langle {\bf p} \rangle \neq {\bf 0} \,.
\label{AM_noncons}    
\end{equation}
This is a consequence of the non-collinear mean group velocity and mean momentum, Fig.~\ref{Fig1}.

On the other hand, using the extrinsic-intrinsic decomposition with the energy centroid, 
\begin{equation}
\langle {\bf L} \rangle^{{\rm ext}\prime} = \langle {\bf r}_E \rangle \times \langle {\bf p} \rangle\,, \quad
\langle {\bf L} \rangle^{{\rm int}\prime} = \langle {\bf L} \rangle - \langle {\bf L} \rangle^{{\rm ext}\prime} \,,
\label{AM_energy}    
\end{equation}
yields conserved extrinsic and intrinsic angular momenta:
\begin{equation}
\frac{d \langle {\bf L} \rangle^{{\rm ext}\prime}}{dt} = \frac{d \langle {\bf r}_E \rangle}{dt} \times \langle {\bf p} \rangle = {\bf 0} \,.
\label{AM_cons}    
\end{equation}
This is a strong argument in favour of the definition (\ref{AM_energy}). Still, it should be emphasized that this property holds true only for the relativistic dispersion $E({\bf p})$.

It should be noticed that the extrinsic-intrinsic decomposition is just a theoretical convention. To the best of our knowledge, these parts do not manifest differently in experiments (in contrast to the spin and orbital parts of the angular momentum). An experiment can only measure the total angular momentum $\langle {\bf L} \rangle$ with respect to some axis determined by the setup.   

\section{Concluding remarks}

We have considered two extensions of the Ehrenfest relation between the mean velocity and momentum of a quantum particle to the relativistic case. First, the velocity of the probability centroid (which is the expectation value of the coordinate operator) can be interpreted as the mean group velocity and it is not proportional to the mean momentum of the particle. Second, the velocity of the energy centroid is proportional to the mean momentum, which is a consequence of the relativistic boost-momentum conservation. Thus, in contrast to non-relativistic particles, the mean group velocity does not generally collinear with the mean momentum. 

We have shown that these features are determined by the dispersion $E({\bf p})$ of the field under consideration. The non-relativistic quadratic dispersion $E\propto p^2$ ensures the proportionality of the mean velocity to the mean momentum (the usual non-relativistic Ehrenfest relation). In turn, the relativistic dispersion $E=\sqrt{p^2c^2 + m^2c^4}$ determines the proportionality of the energy-centroid velocity to the mean momentum. 
For any other dispersions, these properties do not hold true. 

As a consequence, for non-relativistic particles, the intrinsic and extrinsic orbital angular momenta determined with respect to the probability centroid are separately conserved quantities. For relativistic particles, similar angular momenta determined with respect to the energy centroid are conserved in free space. Since this decomposition and conservation depend on the dispersion, it mostly has an illustrative meaning, and only the total angular momentum is a fundamental conserved quantity.  

The difference between the probability and energy centroids of relativistic distributed objects have been known before \cite{Muller1992AJP, Bliokh2012PRL}. A difference between the magnitudes of their velocities was noticed recently \cite{Bliokh2023JPA}, while the difference in their directions has not been described until now. Since the features discussed here are solely determined by the dispersion, these are relevant to any waves, including classical electromagnetic or acoustic waves with massless relativistic-like dispersions. Note, however, that the difference between the probability and energy centroids manifests itself only in non-monochromatic fields (pulses), localized in space and time. This difference and the extrinsic-intrinsic angular momentum decompositions were recently discussed \cite{Bliokh2023PRA} in relation to spatiotemporal vortex pulses \cite{Sukhorukov2005, Bliokh2012PRA, Jhajj2016, Chong2020NP, Bliokh2021PRL, Zhang2023NC, Che2024PRL, Vo2024}. In particular, the non-conservation of the extrinsic and intrinsic angular momenta determined with respect to the probability centroid was noticed in \cite{Porras2024} for a rather nontrivial structured pulse, while for more symmetric pulses with an elliptical wavefunction distribution in the momentum space \cite{Bliokh2023PRA} these angular momenta are conserved, because the probability and energy centroids move along parallel trajectories in the same direction.

We hope that these considerations shed some light on the nontrivial character of: (i) the Ehrenfest theorem, (ii) relations between the mean velocity and momentum, and  (iii) the intrinsic/extrinsic angular momentum decomposition, beyond the simplest Schr\"{o}dinger Hamiltonian.
 
\vspace*{0.1cm}
\begin{acknowledgments}
I acknowledge fruitful discussions with Profs. Aleksandr Y. Bekshaev and Miguel A. Porras, as well as support from Ikerbasque (Basque Foundation of Science), 
Marie Sk\l{}odowska-Curie COFUND Programme of the European Commission (project HORIZON-MSCA-2022-COFUND-101126600-SmartBRAIN3), 
ENSEMBLE3 Project (MAB/2020/14) which is carried out within the International Research Agendas Programme (IRAP) of the Foundation for Polish Science co-financed by the European Union under the European Regional Development Fund and Teaming Horizon 2020 programme (GA. No. 857543) of the European Commission and the project of the Minister of Science and Higher Education ``Support for the activities of Centers of Excellence established in Poland under the Horizon 2020 program'' (contract MEiN/2023/DIR/3797).
\end{acknowledgments}



\bibliography{References_STVP}

\end{document}